\documentclass[journal=jacsat,manuscript=article]{achemso}

\usepackage[version=3]{mhchem} 
\usepackage{svg}
\usepackage{siunitx}      
\usepackage{booktabs}
\usepackage{graphicx}
\usepackage{xcolor} 
\usepackage{tikz}
\usetikzlibrary{shapes.geometric,arrows,calc}

\definecolor{OrangeColor}{HTML}{d4580d} 

\usepackage{amsmath}
\usepackage{amssymb}
\DeclareMathOperator*{\argmax}{arg\,max}

\SectionNumbersOn          
\author{Khayrul Islam}
\affiliation[]{Department of Mechanical Engineering, Lehigh University, Bethlehem, PA 18015, USA}
\alsoaffiliation{Computational Engineering Department, Lawrence Livermore National Laboratory, Livermore, CA 94550, USA}

\author{Mehedi Hasan}
\affiliation[]{Department of Industrial and Production Engineering, Bangladesh University of Engineering and Technology, Dhaka--1000, Bangladesh}

\author{Yaling Liu}
\affiliation[]{Precision Medicine Translational Research Center, West China Hospital, Sichuan University, Chengdu, Sichuan, 610041, China}
\alsoaffiliation{Center for High Altitude Medicine, West China Hospital, Sichuan University, Chengdu, Sichuan, 610041, China}
\alsoaffiliation{Department of Bioengineering, Lehigh University, Bethlehem, PA 18015, USA}
\email{yaling.liu@gmail.com}

\title[An \textsf{achemso} demo]
  {Physics-Guided Surrogate Modeling for Machine Learning–Driven DLD Design Optimization}

\abbreviations{IR,NMR,UV}

\begin{document}

\begin{abstract}
Microfluidic separation technologies have transformed label-free cell sorting by exploiting intrinsic biophysical properties, yet the translation of these platforms from laboratory prototypes to clinical applications remains constrained by the empirical, trial-and-error nature of device design. Deterministic Lateral Displacement (DLD) represents a paradigmatic example: while demonstrating robust discrimination of cells by size, shape, and deformability across diverse applications including circulating tumor cell isolation and malaria diagnostics, DLD performance exhibits extreme sensitivity to the coupled interplay between cellular mechanical phenotype and micron-scale geometric parameters, necessitating iterative fabrication-testing cycles that span weeks to months. We present the first complete inverse design framework that transforms measured cellular deformability into fabrication-ready DLD specifications through physics-guided machine learning. Our approach integrates high-fidelity lattice-Boltzmann and immersed-boundary simulations with gradient-boosted surrogate models to systematically map cellular mechanical properties to migration behavior across manufacturing-feasible geometric configurations (pillar radius, gap, periodicity). Type II ANOVA quantifies the relative influence of these parameters, revealing that while geometric factors dominate migration angle variance ($F = 63.72$, $p < 10^{-37}$), cellular deformability exerts statistically significant effects through interactions with device geometry ($F = 48.23$, $p < 10^{-34}$). The resulting XGBoost surrogate achieves sub-degree predictive accuracy ($R^2 = 0.9999$, MSE $= 2 \times 10^{-4}$), enabling Bayesian optimization via tree-structured Parzen estimation to identify optimal array architectures in under 60 seconds—reducing design iteration from weeks of experimental prototyping to minutes of automated computation. By deploying this validated pipeline as an accessible web application that accepts experimentally measured deformation indices and returns optimized device specifications with tolerance analysis, we democratize DLD design for researchers without specialized computational expertise, thereby accelerating the translation of microfluidic technologies from research-grade prototypes to application-specific, clinically deployable devices.
\end{abstract}

\section{Introduction}

Deterministic Lateral Displacement(DLD) has emerged as a transformative microfluidic technology that enables label-free, passive separation of biological particles based on their intrinsic physical properties. Since its introduction by Huang et al. in 2004\cite{huang2004continuous}, DLD has evolved from a purely size-based sorting mechanism into a versatile platform capable of discriminating particles by shape\cite{zeming2013rotational}, deformability\cite{holmes2014separation}, and electrical properties\cite{beech2009tipping}. This versatility has enabled DLD to address critical challenges across diverse biomedical domains. Label-free microfluidic DLD devices now perform Circulating Tumor Cell (CTC) enrichment for liquid biopsy \cite{lee2018clearcell}, isolate fetal red blood cells for non-invasive prenatal testing \cite{chen2023correction}, and provide label-free malaria diagnostics through detection of pathogen-induced erythrocyte stiffening \cite{kruger2014deformability}. In forensic applications, DLD achieves greater than 90\% male DNA purity in sperm-epithelial cell separation\cite{liu2015separation}, demonstrating regulatory readiness for clinical, industrial, and legal workflows. Recent integration of Field-Programmable Gate Array (FPGA) accelerated deep learning with microfluidic systems has enabled real-time cell classification and sorting at high throughput\cite{Islam2025-ed, Islam2025-dp}, further expanding the capabilities of label-free separation technologies. Unlike conventional cell-sorting approaches that depend on antibody labeling or fluorescent tagging\cite{miltenyi1990high,bonner1972fluorescence,shapiro2005practical}, DLD exploits the innate mechanical phenotype of cells\cite{preira2013passive,wang2015microfluidic}, offering advantages in cost, simplicity, and preservation of cell viability for downstream assays.

Despite these successes, the translation of DLD technology from research-grade prototypes to application-specific devices remains bottlenecked by a fundamental design challenge: the extreme sensitivity of separation performance to the coupled interplay between cellular mechanical properties and device geometry. Device performance is acutely sensitive to micron-scale geometric tolerances and hydrodynamic conditions. Even modest deviations in pillar gap, array inclination, or sample viscosity can shift the critical particle diameter or capillary number sufficiently to nullify the intended separation window\cite{inglis2006critical,fan2025latest}. This sensitivity is compounded by the intrinsic coupling between size- and deformability-dependent migration modes: a compliant cell may behave hydrodynamically as though it were smaller than its physical diameter, complicating the design of arrays capable of resolving heterogeneous populations whose elastic moduli span orders of magnitude\cite{holmes2014separation,kabacaouglu2018optimal}. 

Traditional design approaches attempt to address this complexity through empirical scaling laws derived from rigid-particle experiments and iterative fabrication-testing cycles. For instance, the canonical critical diameter relationship $D_c \approx 1.4\,P_g\,\varepsilon^{0.48}$, where $P_g$ is the pillar gap and $\varepsilon$ is the row-shift fraction\cite{inglis2006critical}, provides a useful first-order estimate for rigid spheres but fundamentally fails to capture the mechanics of deformable cells. This expression cannot predict migration behavior for compliant particles, irregularly shaped cells, or operation at elevated Reynolds or capillary numbers where nonlinear hydrodynamic effects become significant\cite{dincau2018deterministic,mallorie2024numerical,wullenweber2023numerical}. Moreover, these empirical rules offer no guidance on optimizing array geometry (pillar radius, gap width, periodicity) for specific cellular phenotypes characterized by measured deformability indices. Consequently, device development proceeds through costly and time-intensive experimental iteration, with designers manually adjusting geometric parameters based on trial-and-error observations\cite{musharaf2024computational}. For applications requiring separation of closely matched phenotypes or operation near critical conditions, this process can extend across months of fabrication cycles, hindering rapid prototyping and personalized device optimization.

Recent efforts have begun to leverage computational modeling and machine learning to accelerate DLD design. High-fidelity multiphysics simulations that couple Lattice-Boltzmann(LB)fluid solvers with Immersed-Boundary(IB) models of deformable membranes now resolve single-cell migration trajectories, capturing the fluid-structure interactions that govern deformability-based separation~\cite{kruger2014deformability}. However, translating such simulation insights into actionable geometry recommendations remains largely manual and expert-driven. Emerging machine-learning approaches, ranging from mode-classification and regression trained on limited simulation or experimental datasets to early attempts at automated or inverse design, have begun to bridge this gap but remain fragmented and preliminary~\cite{gioe2022deterministic,vatandoust2022dldnn,liu2024unraveling}. Critically, existing frameworks face four limitations that constrain practical deployment. \textit{First}, many models are still forward predictors of migration or critical diameter given a prescribed geometry, with only nascent demonstrations of inverse recommendation~\cite{gioe2022deterministic,vatandoust2022dldnn}. \textit{Second}, several surrogates under-represent deformability as an input, despite evidence that cell mechanics and array geometry interact nonlinearly to determine trajectories~\cite{kabacaouglu2018optimal}. \textit{Third}, robustness and tolerance propagation are rarely incorporated, even though microfabrication variability on the order of microns (e.g., $\pm$2~$\mu$m device height, $\pm$1.5$^\circ$ wall angle) can materially shift separation behavior~\cite{fachin2017monolithic}. \textit{Fourth}, there is still no widely adopted, end-to-end pipeline that couples high-fidelity simulation, statistical validation, inverse optimization, and user-friendly deployment for experimentalists, an oft-noted gap in the DLD community~\cite{hochstetter2020deterministic}.

To address these gaps, this study develops the first complete, physics-guided machine learning framework for inverse DLD design that transforms measured cellular deformability into fabrication-ready device geometry. We combine high-throughput multiphysics simulations coupling LB fluid dynamics with IB membrane mechanics, supervised learning, statistical interaction analysis, and Bayesian optimization to establish a bidirectional mapping between cellular mechanical phenotype and optimal array architecture. Our framework quantifies the relative influence of geometric parameters and deformability through Type II ANOVA, revealing that while pillar gap, radius, and periodicity dominate main effects, deformability exerts its influence primarily through statistically significant interactions with geometry ($p < 10^{-30}$). These insights inform a tolerance-aware inverse design algorithm that identifies robust operational regimes where separation performance is minimally sensitive to manufacturing variability. To maximize accessibility and accelerate adoption, we deploy the complete pipeline as a containerized web application with sub-minute optimization times, enabling researchers to input measured deformability values and immediately obtain optimized, fabrication-ready DLD specifications.

The contributions of this work are threefold:

\begin{enumerate}
    \item \textbf{\texttt{Comprehensive Simulation Pipeline:}} High-fidelity computational framework systematically mapping cellular mechanical properties (stretching, bending, area-dilation moduli) to Deformation Index(DI) and migration angle across physiologically relevant stiffness ranges and manufacturing-feasible geometric configurations (pillar radius, gap, periodicity), enabling robust surrogate model training.
    
    \item \textbf{\texttt{Robust ML Surrogates:}} Statistical framework quantifying deformability-geometry interactions (Type II ANOVA, $F > 35$, $p < 10^{-25}$) coupled with interpretable XGBoost surrogates ($R^2 = 0.9999$, sub-degree error) enabling Bayesian inverse design and robustness region mapping under fabrication tolerances.
    
    \item \textbf{\texttt{Web-Based Deployment:}} Production-ready service accepting measured DI, executing constrained optimization ($<$60 seconds), and returning fabrication-ready specifications with sensitivity analysis, thereby democratizing DLD design for experimentalists.
\end{enumerate}

\section{Methods}
In silico modeling of cell behavior under microfluidic constraints has become an indispensable tool for probing deformation dynamics, migration patterns, and mechanical phenotype classification. Among the available frameworks, hybrid fluid–structure interaction solvers combining LB methods for flow and IB techniques for membrane dynamics have demonstrated exceptional fidelity in capturing the biomechanics of soft particles, including red blood cells and compound capsules\cite{islam2023-vp}. Analogous molecular dynamics frameworks have successfully captured mechanical behavior in complex material systems through systematic variation of structural parameters~\cite{Hasan2024-nr}. Motivated by these advances, we adopted a similar multiscale approach to simulate cell deformation and transport across confined geometries.

\subsection{Lattice-Boltzmann Flow Solver}
\label{sec:LB}

Micron-scale hydrodynamics are resolved with the three-dimensional, nineteen-velocity LB scheme (D3Q19-LB) available in \textsc{ESPResSo}. Rigid channel walls and every cylindrical pillar are treated as \emph{mid-plane bounce-back} boundaries, a choice that recovers second-order-accurate no-slip conditions without explicit wall meshing. Unless stated otherwise, both plasma and cytosol share identical lattice density and kinematic viscosity, $\rho = 1$ and $\nu = 1.5$, and are driven by a homogeneous body-force density $f_x$ that mimics the axial pressure gradient of DLD microfluidics.

Fluid nodes are kinematically linked to the deformable membranes via a force- and torque-free IB dissipative-coupling (IB-DC) algorithm. With a friction coefficient $\gamma=1.56$ lattice units, the procedure (i) interpolates local LB velocities to each membrane vertex, (ii) evaluates membrane forces \textit{in situ}, (iii) redistributes those forces to the surrounding fluid nodes, and (iv) dissipates the momentum exchange locally. The result is a self-consistent lubrication layer that persists even under the extreme confinements encountered in sub-gap DLD operation. Details of the periodic unit cell strategy and computational efficiency are provided in Section~\ref{sec:pipeline}.

\subsection{Immersed-Boundary Model for Cell Mechanics}
\label{sec:IB}

Each cell membrane is discretised as a quasi-equidistant triangular network ($\sim$2000 vertices, $\sim$2000 facets) generated from the analytical Evans-Fung profile and rescaled isotropically to a radius of $7.5$ lattice units. The elastic energy of the network reads

\begin{equation}
\label{eq:membrane_energy}
\begin{aligned}
W &= \frac{k_s}{2}\,(L-L_0)^2
+ \frac{k_b}{2}\,(\theta-\theta_0)^2
+ \frac{k_\alpha}{2}\!\left(\frac{A-A_0}{A_0}\right)^{\!2}
\\
&\quad
+ \frac{k_{ag}}{2}\!\left(\frac{S-S_0}{S_0}\right)^{\!2}
+ \frac{k_v}{2}\!\left(\frac{V-V_0}{V_0}\right)^{\!2},
\end{aligned}
\end{equation}

\noindent where the five terms penalise (i) Neo-Hookean stretching ($k_s$), (ii) Helfrich-style bending ($k_b$), (iii) local-area variation ($k_\alpha$), (iv) global-area change ($k_{ag}$), and (v) enclosed volume change ($k_v$). The continuum surface-compressibility modulus follows as
\begin{equation}
K = \tfrac32 k_s + \tfrac12 k_\alpha.
\label{eq:K}
\end{equation}

All five force contributions are formulated such that the net force and torque on an isolated membrane patch vanish identically at every time step. This guarantees translational and rotational invariance without auxiliary constraints, a prerequisite for unbiased hydrodynamic coupling. Table~\ref{tab:membrane_moduli} lists the five stiffness sets explored; the effective surface-compression modulus covers $K\approx 4$--$25~\mathrm{mN\,m^{-1}}$, bracketing the experimentally reported range for human erythrocytes and enabling a controlled sensitivity analysis. Surface viscosity is incorporated via a Kelvin-Voigt dashpot of magnitude $k_{\text{visc}}=1.5~\si{\micro\newton\second\per\metre}$, rendering the membrane \emph{dissipative yet fully reversible}. Membrane vertices advance with a velocity-Verlet algorithm that shares the LB time step $\Delta t=0.1$.

\begin{table}[ht]
\centering
\setlength{\tabcolsep}{3pt}
\begin{tabular}{@{}llccccc@{}}
\toprule
\textbf{Contribution} & \textbf{Symbol} & \textbf{Set 1} & \textbf{Set 2} & \textbf{Set 3} & \textbf{Set 4} & \textbf{Set 5} \\
\midrule
Stretching   & $k_s$   & 0.00510 & 0.00760 & 0.01524 & 0.01778 & 0.02030 \\
Bending      & $k_b$   & $4.3 \times 10^{-5}$ & $9.5 \times 10^{-5}$ & $2.50 \times 10^{-4}$ & $3.00 \times 10^{-4}$ & $3.50 \times 10^{-4}$ \\
Local area   & $k_{\alpha}$& 0.0102 & 0.0152 & 0.0300 & 0.0350 & 0.0406 \\
Global area  & $k_{ag}$ & 0.50 & 0.50 & 0.50 & 0.50 & 0.50 \\
Volume       & $k_v$    & 0.90 & 0.90 & 0.90 & 0.90 & 0.90 \\
Surface viscosity & $k_{\mathrm{visc}}$ & 1.5  & 1.5  & 1.5  & 1.5  & 1.5  \\
\bottomrule
\end{tabular}
\vspace{-.2cm}
\caption{Elastic coefficients explored in the parameter sweep (\si{\pico\newton\per\micro\metre} for $k_s$, $k_{\alpha}$; \si{\pico\newton\micro\metre} for $k_b$).}
\label{tab:membrane_moduli}
\end{table}

\subsection{Periodic Unit Cell and Simulation Pipeline}
\label{sec:pipeline}

\begin{figure}[htbp]
  \centering
  \includegraphics[width=0.9\textwidth,page=1]{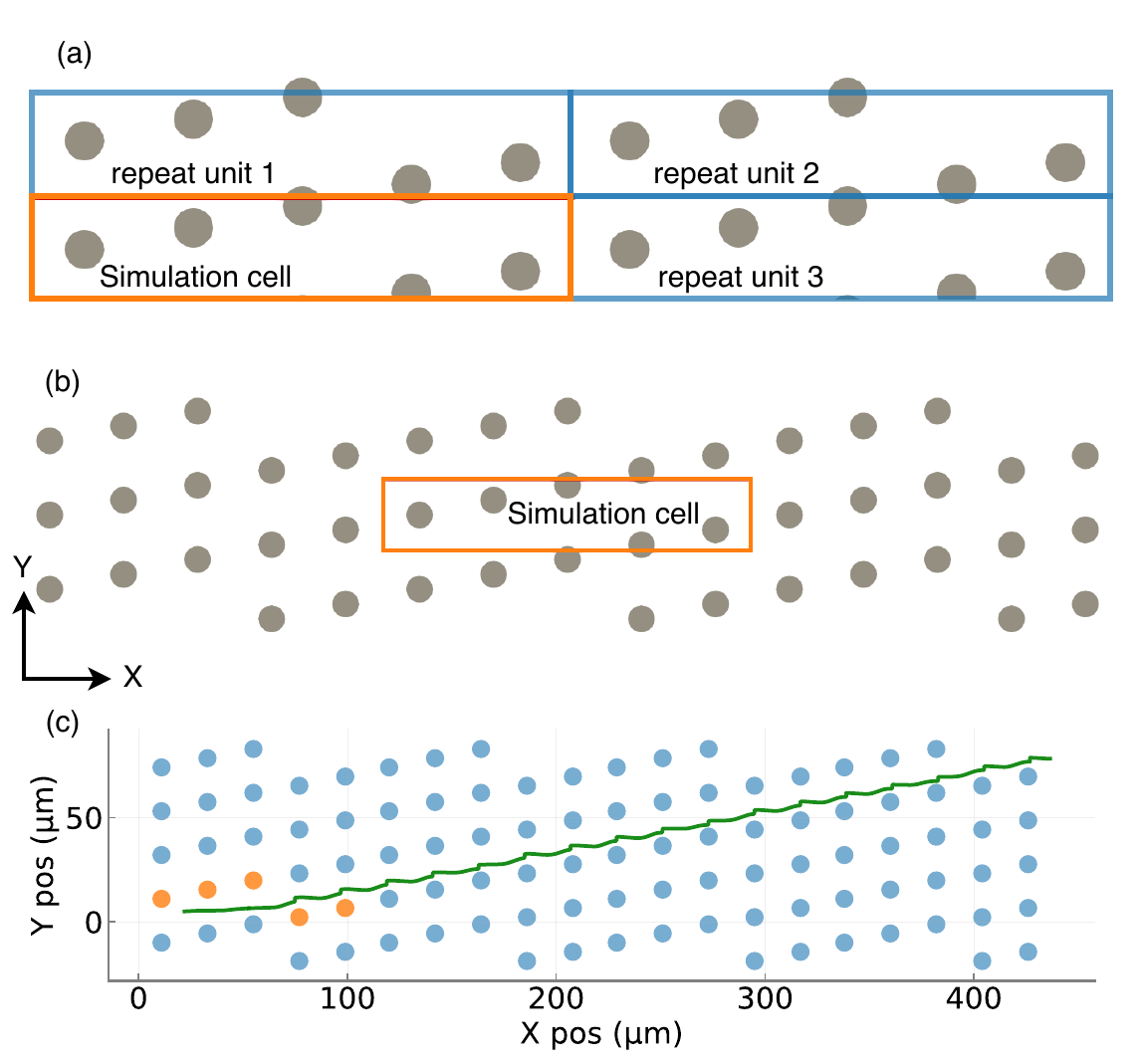}
\caption{Periodic unit-cell simulation strategy. (a) Computational domain (orange) with periodic repeat units (blue) containing pillar array (gray circles), (b) simulation cell context within full lattice, and (c) cell trajectory: pillars (blue), cell positions (orange), migration path (green). Migration angle extracted from displacement slope.}
\label{fig:periodic_domain}
\end{figure}

The predictive framework developed in this study requires systematic exploration of the design space spanned by cellular mechanical properties and device geometry. To achieve this while maintaining computational efficiency, we simulate a single \emph{periodic unit cell} of the DLD lattice rather than an entire device. Figure~\ref{fig:periodic_domain} illustrates this computational strategy. The periodic unit cell (Fig.~\ref{fig:periodic_domain}a, orange region) contains the minimal geometric configuration necessary to capture one complete DLD period, with pillar arrays (gray circles) arranged according to the specified row shift. This cell is spatially replicated through periodic boundary conditions (blue regions) to reproduce the infinite lattice structure. As shown in Fig.~\ref{fig:periodic_domain}b, the simulation domain represents a single repeating unit embedded within the full device context, thereby eliminating the computational burden of explicitly modeling hundreds of pillar columns while retaining all essential flow features: wake recirculation, near-wall velocity gradients, and cell-pillar interactions.


The total number and placement of pillars are determined by the periodicity parameter $N_p$, which defines the lattice shift per row. For each parameter triplet $(P_r, P_g, N_p)$ sampled during design space exploration, the domain dimensions are set as $d_c = 2P_r + P_g$ (pillar center-to-center spacing), $L_x = N_p \cdot d_c$ (streamwise length), and $L_y = d_c$ (transverse length). Periodic boundary conditions are imposed in both axial ($x$) and transverse ($y$) directions, with the row shift explicitly encoded as $d = L_y / N_p$. Because $d$ is constrained to be an integer multiple of the lattice spacing, no interpolation of distribution functions is required at the boundaries, preserving the second-order accuracy of the D3Q19 scheme. This approach reduces computational volume by approximately two orders of magnitude compared to full-device models: typical experimental devices contain $\sim$100 rows spanning $\sim$10 periods (volume $V_{\text{full}} \approx 1000 \, d_c^2 h$), whereas the unit cell simulates only $N_p \in [4, 12]$ rows in one period (volume $V_{\text{unit}} = N_p \cdot d_c^2 h$), yielding a reduction factor of $\sim$80--250.

Pillar surfaces are constructed analytically and coupled to both the fluid solver and membrane collision detection system. A short-range soft-sphere repulsion potential, $V(r) = a\,r^{-n}$ for $r < r_c$ with $a = 10^{-4}$ and $n = 1.2$, enforces minimum clearance between membrane vertices and rigid obstacles without introducing artificial stiffening of the hydrodynamic coupling. This repulsive interaction preserves physically realistic lubrication layers in the confined gaps characteristic of sub-critical DLD operation. Cell migration behavior is quantified through post-processing that exploits the periodic structure to extract unbiased statistics. The trajectory analysis algorithm identifies entrance and exit checkpoints separated by exactly one complete lattice period, corresponding to a streamwise displacement of $L_x = N_p \cdot d_c$ and a lateral shift of $d = L_y/N_p$, thereby eliminating artifacts from partial traversals. Figure~\ref{fig:periodic_domain}c visualizes a representative trajectory, where the cell (orange markers) navigates through the pillar array (blue circles) following a characteristic migration path (green line). The net displacement vector $(\Delta x, \Delta y)$ between these checkpoints yields the migration angle
\[
\theta_m = \arctan\left(\frac{\Delta y}{\Delta x}\right) \times \frac{180}{\pi},
\]
which serves as the primary response variable for surrogate model training. This scalar metric enables direct comparison across the full parameter space while naturally averaging out local fluctuations arising from membrane elasticity and transient flow perturbations.

\subsection{Validation of Simulation with Experimental Data}

To establish the predictive accuracy of the computational framework, we performed quantitative validation against experimental measurements of cell deformation through microfluidic constrictions. Table~\ref{tab:validation} presents a direct comparison between experimentally measured and numerically predicted cell dimensions at four characteristic positions along the flow channel. The horizontal diameter ($D$) and vertical diameter ($H$) were extracted from both experimental microscopy images and simulation snapshots using consistent image analysis protocols to ensure methodological parity.

\begin{table}[ht]
\centering
\setlength{\tabcolsep}{3pt}
\begin{tabular}{@{}ccccccc@{}}
\toprule
\textbf{Position} & \multicolumn{2}{c}{\textbf{Experiment}} & \multicolumn{2}{c}{\textbf{Simulation}} & \multicolumn{2}{c}{\textbf{Deviation (\%)}} \\
\cmidrule(lr){2-3}\cmidrule(lr){4-5}\cmidrule(lr){6-7}
 & $D$ ($\mu$m) & $H$ ($\mu$m) & $D$ ($\mu$m) & $H$ ($\mu$m) & $D$ & $H$ \\
\midrule
1 & 19.734 & 16.144 & 20.145 & 17.038 & 2.08 & 5.54 \\
2 & 25.147 & 13.472 & 24.368 & 13.844 & 3.10 & 2.76 \\
3 & 32.565 & 5.004  & 32.384 & 4.982  & 0.56 & 0.44 \\
4 & 19.764 & 11.794 & 18.934 & 10.468 & 4.20 & 11.25 \\
\bottomrule
\end{tabular}
\vspace{0.2cm}
\caption{Quantitative comparison between experimental and simulated horizontal ($D$) and vertical ($H$) cell diameters across four characteristic positions in the constriction channel. Deviations represent absolute percentage differences between simulation and experiment.}
\label{tab:validation}
\end{table}

The validation dataset encompasses four distinct deformation regimes: initial entry (Position 1), progressive elongation (Position 2), maximum confinement at the channel throat (Position 3), and downstream relaxation (Position 4). At Position 1, the cell undergoes initial lateral stretching as it encounters the constriction entrance, with simulated diameters deviating by 2.08\% ($D$) and 5.54\% ($H$) from experimental measurements. Position 2 captures the transition regime where axial compression intensifies, yielding deviations of 3.10\% and 2.76\%, respectively. Most notably, Position 3—corresponding to maximum deformation at the throat where membrane mechanics and hydrodynamic stresses reach equilibrium—exhibits exceptional agreement with errors below 1\% for both dimensions (0.56\% for $D$, 0.44\% for $H$). This sub-percent accuracy at peak deformation validates the fidelity of both the constitutive membrane model and the fluid-structure coupling algorithm under extreme confinement conditions.

Position 4, representing the recovery phase as the cell exits the constriction, shows somewhat elevated deviations (4.20\% for $D$, 11.25\% for $H$). The higher vertical discrepancy is attributed to two factors: first, the dynamic nature of viscoelastic relaxation during this transient phase makes diameter measurements highly sensitive to the precise instant of observation, which differs slightly between experiment (captured at finite frame rates) and simulation (computed at fixed time intervals); second, experimental contour extraction during rapid shape recovery introduces greater uncertainty due to reduced image contrast as the cell membrane reorients. Importantly, the horizontal diameter deviation remains below 5\% even in this challenging regime, confirming robust predictive capability across all flow phases.

The mean absolute deviation across all positions is 2.49\% for $D$ and 4.99\% for $H$, both well within the acceptable bounds for cellular-scale biomechanical simulations and substantially lower than typical experimental measurement uncertainties (5--10\%) in microfluidic cell deformability assays\cite{islam2023-vp}. Furthermore, the simulation correctly captures the qualitative deformation sequence: lateral stretching upon entry, progressive axial elongation through the constriction, maximal deformation at the throat, and downstream shape recovery. This morphological fidelity, combined with quantitative diameter agreement, demonstrates that the lattice-Boltzmann fluid solver, immersed-boundary membrane coupling, and elastic energy formulation (Eq.~\ref{eq:membrane_energy}) collectively reproduce the essential physics governing cell-scale fluid-structure interactions.

The successful validation against independent experimental data establishes confidence in the computational framework's ability to generate physically realistic training datasets for surrogate model development. Since the machine learning surrogates presented in subsequent sections are trained exclusively on simulation data, this validation provides critical assurance that the predicted geometry-deformability-migration relationships reflect true biophysical behavior rather than numerical artifacts. The demonstrated sub-5\% accuracy for horizontal dimensions (the primary determinant of lateral migration in DLD devices) specifically validates the framework's suitability for predicting cell trajectories across the geometric parameter space explored in this study.
\section{Results and discussion}

\subsection{Deformation Index Sensitivity to Membrane Mechanics}
\label{sec:DI_mech}

\begin{figure}[htbp]
  \centering
  \includegraphics[width=0.9\textwidth,page=1]{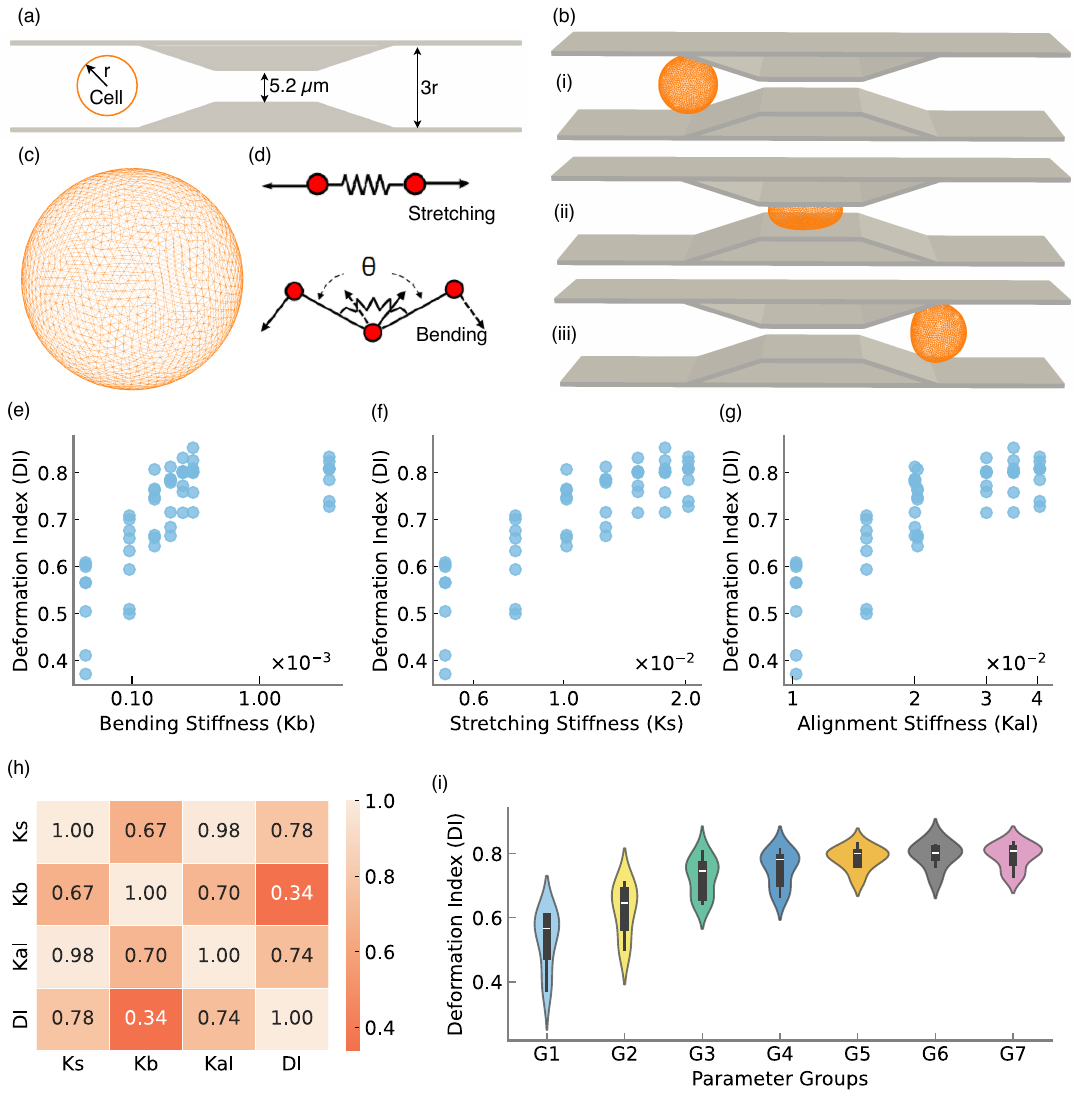}
\caption{Computational assessment of membrane mechanical parameters influencing cellular deformation. (a) Microfluidic constriction geometry, (b) representative deformation stages: (i) entry, (ii) midpoint, and (iii) exit, (c--d) membrane model with stretching and bending springs, (e--g) scatter plots of DI versus $K_s$, $K_{\alpha}$, and $K_b$, showing strong mechanical dependence with minor positional scatter, (h) Pearson correlation heatmap quantifying parameter interdependence, and (i) violin plot demonstrating convergence of DI distributions with increasing membrane stiffness across seven mechanical groups.}
  \label{fig:cell_sq}
\end{figure}

The DI serves as a widely adopted quantitative descriptor of cellular mechanical phenotype in microfluidic assays, particularly for high-throughput characterization of cell populations under physiologically relevant flow conditions\cite{otto2015real,fregin2019high}. The metric is defined as

\begin{equation}
\mathrm{DI} = \frac{a - b}{a + b},
\label{eq:DI}
\end{equation}

where $a$ and $b$ denote the major and minor axes of the cell's projected profile in the imaging plane, respectively, extracted from bounding box analysis. While DI fundamentally reflects the intrinsic mechanical properties of the cell membrane and cytoskeleton, experimental measurements exhibit inherent variability arising from three principal sources: biological heterogeneity within cell populations, measurement uncertainties in image-based feature extraction, and sensitivity to initial positioning within the flow field\cite{fregin2019high,mokbel2017numerical}. 

Recent investigations of microfluidic deformability cytometry have systematically characterized these sources of measurement variation\cite{fregin2019high}. Positional offsets from the flow centerline, velocity fluctuations in inlet channels, and variations in longitudinal seeding position can introduce scatter in DI measurements even for cells with identical mechanical properties. However, comparative studies across multiple high-throughput platforms demonstrate that when cells are properly focused along the channel centerline and subjected to controlled hydrodynamic conditions, the dominant contribution to DI variability arises from true biological variation in cellular mechanical properties rather than experimental artifacts\cite{otto2015real,urbanska2020physics}. Cross-laboratory validation studies comparing constriction-based, shear-based, and extensional flow deformability cytometry methods confirm that all platforms reliably detect mechanically induced changes in cell populations despite platform-specific measurement uncertainties, establishing DI as a robust mechanical biomarker when appropriate experimental controls are implemented\cite{urbanska2020physics}.

\begin{table}[ht]
\centering
\setlength{\tabcolsep}{3pt}
\begin{tabular}{@{}lcccc@{}}
\toprule
\textbf{Group} & \textbf{$K_s$} & \textbf{$K_b$} & \textbf{$K_{\alpha}$} & \textbf{$v$} \\
\midrule
G1 & 0.0051   & $4.33 \times 10^{-5}$ & 0.0102 & 1.095 \\
G2 & 0.0076   & $9.526 \times 10^{-5}$ & 0.0152 & 1.5495 \\
G3 & 0.01016  & $1.5 \times 10^{-4}$   & 0.0203 & 1.8975 \\
G4 & 0.0127   & $2.0 \times 10^{-4}$   & 0.0200 & 2.121 \\
G5 & 0.01524  & $2.5 \times 10^{-4}$   & 0.0300 & 2.739 \\
G6 & 0.01778  & $3.0 \times 10^{-4}$   & 0.0350 & 3.000 \\
G7 & 0.0203   & $3.5 \times 10^{-3}$   & 0.0406 & 3.0975 \\
\bottomrule
\end{tabular}
\vspace{0.2cm}
\caption{Membrane mechanical parameter sets defining seven groups in the deformation analysis. Each group maintains fixed elastic moduli while varying initial longitudinal position to isolate positional effects.}
\label{tab:group_mappings}
\end{table}

To quantify the relative contributions of membrane mechanics versus positional variability to DI, we conducted a systematic computational analysis in which cells were initialized along the central transverse axis of the constriction while varying only the longitudinal entry coordinate. This configuration isolates the influence of streamwise position on deformation metrics while maintaining consistent lateral alignment. Despite identical membrane properties within each group, including stretching modulus $K_s$, bending modulus $K_b$, and area-dilation modulus $K_{\alpha}$, the resulting DI values exhibit measurable but bounded scatter attributable to spatial variations in local velocity gradients and extensional stress distributions encountered during transit. 

Figure~\ref{fig:cell_sq}(e--g) demonstrates that DI increases monotonically with both $K_s$ and $K_{\alpha}$, exhibiting strongly nonlinear mechanical sensitivity across the physiologically relevant stiffness range. For stretching modulus, DI spans approximately 0.37 to 0.85 across the parameter space, with stiffer membranes resisting axial compression more effectively and consequently achieving greater elongation ratios under equivalent hydrodynamic loading. The dependence on area-dilation modulus $K_{\alpha}$ is most pronounced at lower values between 0.0102 and 0.020, beyond which the response saturates, indicating diminishing marginal sensitivity to further increases in membrane incompressibility. In contrast, bending modulus $K_b$ exhibits threshold behavior: DI increases sharply only when $K_b$ falls below approximately $5 \times 10^{-4}$, then rapidly plateaus, confirming that bending rigidity influences deformation primarily in the low-stiffness regime where membrane flexural modes become energetically accessible.

The Pearson correlation analysis in Figure~\ref{fig:cell_sq}(h) quantitatively corroborates these mechanical dependencies. Both $K_s$ and $K_{\alpha}$ exhibit strong positive correlations with DI, with correlation coefficients of 0.77 and 0.73, respectively, whereas $K_b$ demonstrates only modest correlation at 0.33, consistent with its secondary role under the imposed deformation conditions. The near-unity correlation between $K_s$ and $K_{\alpha}$ of 0.98 highlights their synergistic contribution to resisting in-plane membrane deformation, collectively governing the surface compressibility modulus defined in Eq.~\ref{eq:K}. 

Critically, Figure~\ref{fig:cell_sq}(i) reveals that the magnitude of positional scatter in DI measurements decreases systematically with increasing membrane stiffness. The violin plot distributions for seven mechanical groups spanning G1 through G7 demonstrate progressively narrower spreads despite concurrent increases in flow velocity that would nominally amplify hydrodynamic sensitivity. Table~\ref{tab:group_mappings} details the mechanical parameters for each group. This convergence phenomenon occurs because stiffer membranes impose stronger elastic constraints on achievable deformations, effectively suppressing positional variability and rendering DI increasingly robust to local flow perturbations. In the low-stiffness regime represented by groups G1 through G3, cells remain highly responsive to subtle spatial variations in shear and extensional stresses arising from initial positioning, yielding broader DI distributions with coefficients of variation approaching 15 to 20 percent. Conversely, at higher stiffness levels representative of pathological phenotypes in groups G4 through G7, elastic resistance dominates the mechanical response, constraining DI variability to below 5 to 8 percent despite identical positional randomization protocols.

This stiffness-dependent suppression of positional artifacts directly validates DI as a reliable mechanical input for DLD device optimization. In clinically relevant parameter regimes where mechanical contrast between healthy and pathological phenotypes is maximal, with modulus ratios typically exceeding 2 to 5-fold\cite{swaminathan2011mechanical}, positional scatter of 3 to 5 percent remains substantially smaller than both biological variation of 10 to 15 percent\cite{urbanska2020physics} and the target mechanical contrast that DLD separations are designed to resolve. These results establish that DI is governed primarily by in-plane elastic moduli $K_s$ and $K_{\alpha}$, with bending stiffness contributing only in highly compliant regimes, thereby supporting its deployment as the primary cellular mechanical descriptor in the machine learning framework developed in subsequent sections.

\subsection{Migration Angle Dependence on Cellular and Geometric Parameters}

\begin{figure}[htbp]
  \centering
  \includegraphics[width=\textwidth,page=1]{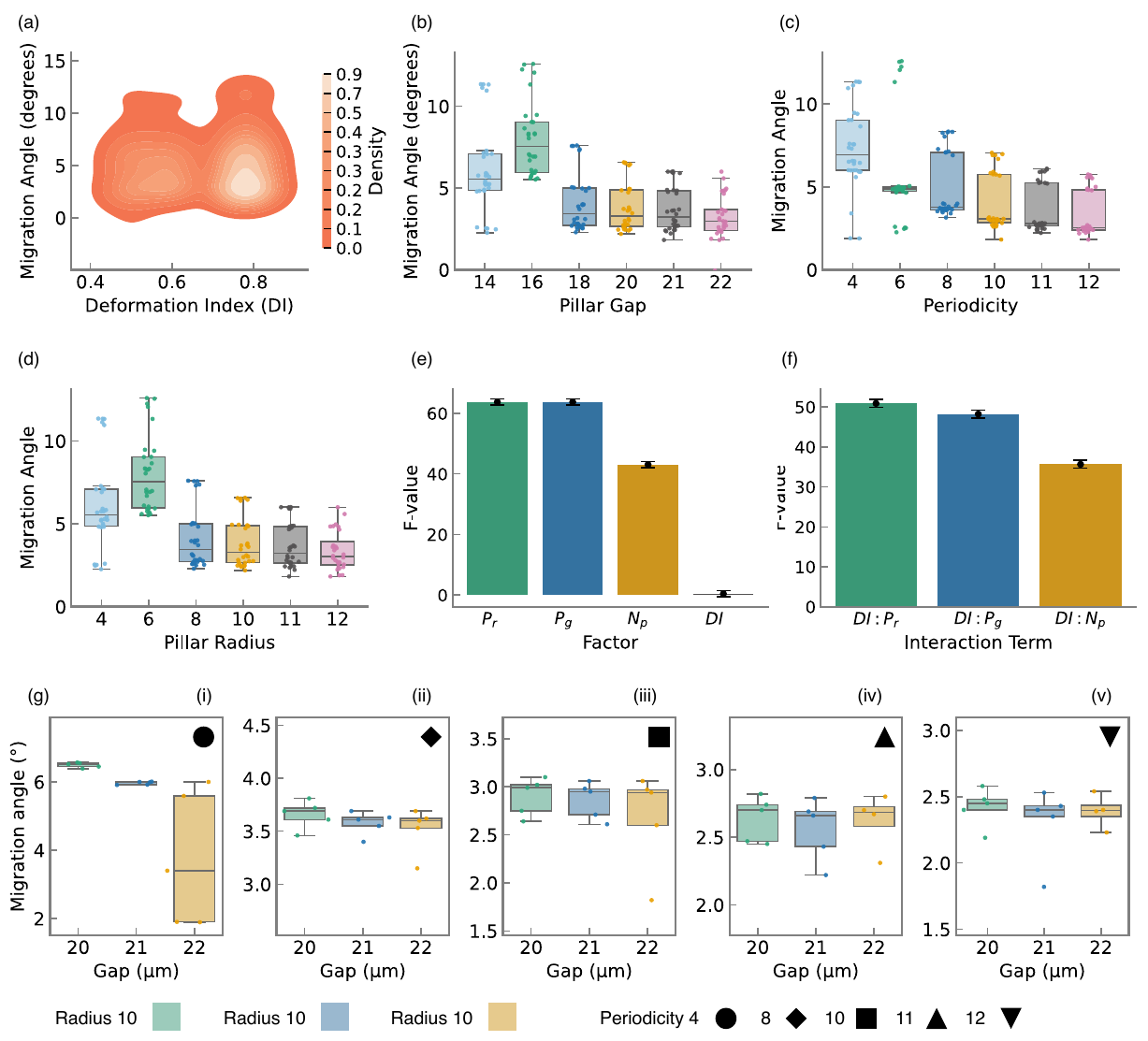}
  \caption{Statistical characterization of migration angle in DLD devices. (a) Kernel density estimate (KDE) of the relationship between pre-characterized DI and migration angle, (b–d) box plots showing migration angle as a function of pillar gap ($P_g$), periodicity ($N_p$), and pillar radius ($P_r$), (e–f) Type II ANOVA results highlighting dominant geometric effects and significant DI-geometry interactions, (g) combined geometric factor analysis using multi-parameter grouping, and (i–v) box plots of migration angle across pillar gaps ($P_g$), where box color encodes pillar radius ($P_r$) and marker shape encodes periodicity ($N_p$).}

  \label{fig:stat_visu}
\end{figure}

Building on the DI characterization in section \ref{sec:DI_mech}, we next examined how mechanical phenotype translates into migration behavior within DLD arrays. Figure~\ref{fig:stat_visu}(a–d) summarizes these relationships. The kernel density estimate in Fig.~\ref{fig:stat_visu}a demonstrates a nonlinear association between DI and migration angle, consistent with prior observations that pre-characterized deformability can predict transport outcomes in microfluidic flows \cite{yamashita2012role,secomb2007two}. Cells with higher DI predominantly migrate at lower angles, while lower DI display larger angular deviations. The emergence of multimodal density regions suggests that deformability interacts with device geometry rather than acting as a sole determinant, in agreement with recent computational frameworks showing that cell mechanics alone provide limited predictive power without hydrodynamic context \cite{cimrak2018computational}.

Geometric parameters exert a dominant influence on migration behavior (Figs.\ref{fig:stat_visu}b–d). Increasing the pillar gap ($P_g$) reduces both the median migration angle and its variability, implying that wider gaps diminish sensitivity to deformability by homogenizing local streamline structures \cite{salafi2019review}. Periodicity ($N_p$), defined as the number of pillar rows per offset cycle, strongly modulates dispersion: low $N_p$ values (4–6) produce broad distributions with elevated migration angles, whereas larger $N_p$ progressively dampens angular excursions, yielding more uniform displacement dynamics. Similarly, pillar radius ($P_r$) controls the degree of lateral redirection. Small radii (4–6µm) enable stronger angular shifts, whereas larger radii (10–12~µm) constrain trajectories to near-axial paths, reducing lateral separation. These findings align with prior studies showing that DLD performance is highly sensitive to pillar geometry and array design \cite{salafi2019review}. Taken together, these results demonstrate that DI provides a compact descriptor of mechanical phenotype, but its predictive value is realized only in the geometric context of the DLD device. Array parameters such as $P_g$, $N_p$, and $P_r$ establish the hydrodynamic envelope within which deformability manifests, underscoring the need to jointly optimize cellular mechanics and device design for effective separation.

\subsubsection*{ANOVA-Based Quantification of Main and Interaction Effects}

To systematically disentangle the contributions of cellular and geometric determinants, we employed a Type~II ANOVA framework with migration angle as the response variable. Explanatory factors included the DI as a surrogate of cellular mechanics, together with the three principal design parameters of the DLD array: pillar gap ($P_g$), pillar radius ($P_r$), and periodicity ($N_p$). The main-effects analysis (Fig.~\ref{fig:stat_visu}e, Table~\ref{tab:anova_factors}) revealed that $P_g$ and $P_r$ were the dominant drivers of migration angle variance (F = 63.72, $p \approx 5.0\times10^{-37}$), with $N_p$ also contributing significantly (F = 43.07, $p \approx 2.0\times10^{-28}$). By contrast, DI alone was not significant (F = 0.40, $p = 0.53$), underscoring that mechanical phenotype in isolation is insufficient to predict migration behavior in a strictly linear fashion. In practical terms, the high F-values and vanishingly small $p$-values for $P_g$, $P_r$, and $N_p$ indicate that these geometric factors explain the overwhelming majority of observed variance in migration angle, whereas the near-zero sum of squares for DI confirms its negligible independent contribution. These findings are consistent with prior reports that device-scale geometry, rather than intrinsic cellular stiffness, establishes the baseline transport behavior in DLD arrays \cite{salafi2019review, cimrak2018computational}.

When interaction terms were incorporated, however, DI emerged as a significant modifier of geometric effects (Fig.~\ref{fig:stat_visu}f, Table~\ref{tab:anova_interactions}). DI–$P_g$ interactions (F = 48.23, $p \approx 4.3\times10^{-34}$), DI–$P_r$ interactions (F = 50.87, $p \approx 6.6\times10^{-32}$), and DI–$N_p$ interactions (F = 35.64, $p \approx 9.3\times10^{-25}$) were all highly significant. Here, the large sum of squares associated with each interaction term indicates that a substantial fraction of migration angle variability arises only when DI is considered jointly with geometry. This finding highlights that DI acts not as a standalone determinant but as a contextual amplifier of geometric confinement. For example, a high-DI (soft) cell exhibits markedly different trajectories when encountering narrow versus wide gaps or small versus large pillars, underscoring the synergistic nature of mechanics–geometry coupling. Such interaction-driven effects have also been reported in computational and experimental frameworks, where deformability-mediated separation was realized only under specific array geometries \cite{ yamashita2012role}.

In sum, the ANOVA results establish that geometry provides the dominant baseline for migration angle determination, but cellular deformability modulates these effects through statistically robust interactions. These insights reinforce the necessity of context-aware design strategies that jointly optimize cellular mechanics and device architecture. Moreover, they provide the analytical foundation for the following section on \textit{Geometric Robustness Region via Operational Stability Mapping}, where we map the parametric domains that sustain stable separation performance against variations in both DI and array design.

\begin{table}[h]
\centering
\caption{Type II ANOVA results for factor influence on migration angle $\theta_m$}
\begin{tabular}{lcccc}
\hline
\textbf{Factor} & \textbf{Sum Sq} & \textbf{df} & \textbf{F-value} & \textbf{$p$-value} \\
\hline
Pillar gap ($P_g$)       & 531.48 & 5 & 63.72 & $5.04 \times 10^{-37}$ \\
Pillar radius ($P_r$)    & 531.48 & 5 & 63.72 & $5.04 \times 10^{-37}$ \\
Periodicity ($N_p$)      & 359.20 & 5 & 43.07 & $2.03 \times 10^{-28}$ \\
Deformation Index ($\mathrm{DI}$)   & 0.67   & 1 & 0.40  & 0.526 \\
\hline
\end{tabular}
\label{tab:anova_factors}
\end{table}

\begin{table}[h]
\centering
\caption{Type II ANOVA results for interaction effects between $\mathrm{DI}$ and DLD geometry factors on $\theta_m$}
\begin{tabular}{lcccc}
\hline
\textbf{Interaction Term} & \textbf{Sum Sq} & \textbf{df} & \textbf{F-value} & \textbf{$p$-value} \\
\hline
$\mathrm{DI} \times P_r$  & 482.67 & 5 & 50.87 & $6.58 \times 10^{-32}$ \\
$\mathrm{DI} \times P_g$  & 549.14 & 6 & 48.23 & $4.32 \times 10^{-34}$ \\
$\mathrm{DI} \times N_p$  & 338.18 & 5 & 35.64 & $9.33 \times 10^{-25}$ \\
\hline
\end{tabular}
\label{tab:anova_interactions}
\end{table}

\subsubsection*{Geometric Robustness Region via Operational Stability Mapping}

To ensure reliable microfluidic performance under manufacturing or biological variability, we sought to determine geometric configurations—defined by pillar radius ($P_r$), gap ($P_g$), and periodicity ($N_p$)—for which the migration angle of cells remains stable. We approached this task from an operational research standpoint by formulating a local stability criterion grounded in neighborhood perturbation theory. Let $\theta(P_r, P_g, N_p)$ denote the observed migration angle for a given triplet of geometry parameters. For each configuration $(P_r^0, P_g^0, N_p^0)$, we defined a local neighborhood $\mathcal{N}$ by varying one geometric parameter at a time while keeping the other two fixed. The corresponding local stability score $S(P_r^0, P_g^0, N_p^0)$ was then computed as the average squared deviation from its neighbors:
\begin{equation}
S(P_r^0, P_g^0, N_p^0) = \frac{1}{|\mathcal{N}|} \sum_{(P_r, P_g, N_p) \in \mathcal{N}} \left[\theta(P_r, P_g, N_p) - \theta(P_r^0, P_g^0, N_p^0)\right]^2.
\end{equation}
Lower values of $S$ indicate geometric setups where the migration angle is less sensitive to perturbations, implying greater operational robustness. We computed $S$ for all 216 unique combinations in our dataset and selected the top $15\%$ most stable configurations as candidates for robust operational zones. Within this subset, we identified the most frequently occurring values of $P_r$, $P_g$, and $N_p$. The resulting robustness ranges were $P_r \in \{10, 11, 12\}$, $P_g \in \{20, 21, 22\}$, and $N_p \in \{4, 8, 10, 11, 12\}$. These values define a stability envelope where the migration angle $\theta_m$ is minimally affected by small local geometric changes.

To visualize these findings, we leveraged the multi-parameter box plots in Fig.~\ref{fig:stat_visu}(g, i–v), where box color encodes pillar radius ($P_r$), marker shape encodes periodicity ($N_p$), and the x-axis represents pillar gap ($P_g$). The narrow spread of most boxes confirms that migration angles exhibit low variance across robust configurations. In parallel, the corresponding strip-plot overlay (markers by $N_p$, colors by $P_r$) reveals a smooth alignment across radii and gaps, further validating the geometric stability of this region. Overall, this section provides a quantitative operational framework for identifying reliable device configurations using migration angle variance as a decision criterion. By maintaining a consistent symbol and color scheme across sections, the mapping of robustness domains remains directly comparable to the preceding statistical analyses, thereby offering an interpretable and deployable pathway for robust DLD device design.

\begin{table}[h]
\centering
\caption{Model performance comparison on the test dataset}
\begin{tabular}{lcc}
\hline
\textbf{Model} & \textbf{Test MSE} & \textbf{Test $R^2$} \\
\hline
XGBoost & 0.0002 & 0.9999 \\
SVR (with Feature Selection) & 0.0003 & 0.9999 \\
Random Forest & 0.0166 & 0.9973 \\
Polynomial Ridge & 0.3159 & 0.9492 \\
Elastic Net & 0.6159 & 0.8804 \\
Linear Regression & 0.6339 & 0.8981 \\
\hline
\end{tabular}
\label{tab:model_comparison}
\end{table}

\begin{figure}[htbp]
  \centering
  \includegraphics[width=\textwidth,page=1]{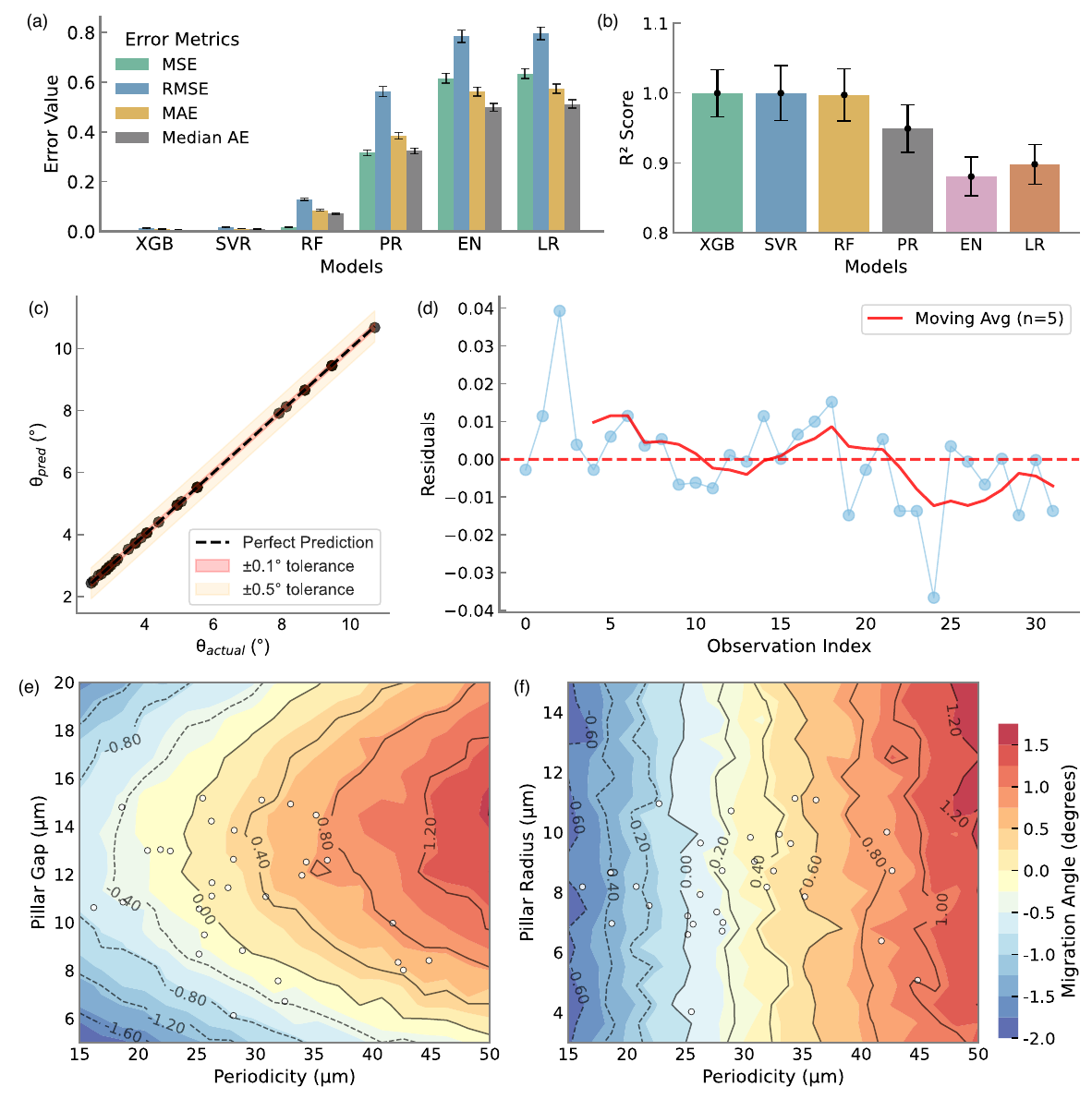}
  \caption{Evaluation and interpretability of machine learning models for migration-angle prediction in DLD devices. (a–b) comparative error metrics and $R^2$ scores across six models (XGBoost, SVR, RF, PR, EN, LR), (c) parity plot of predicted versus actual migration angles using XGBoost, with tolerance bands and ideal $y = x$ line, (d) residual analysis against observation index, showing independence and error trends, and (e–f) feature interaction maps highlighting coupled effects of periodicity with pillar gap and pillar radius on migration angle predictions.}
  \label{fig:model_visu}
\end{figure}

\subsection{Supervised Learning for Forward Prediction in DLD}

Motivated by the preceding analyses—which established that array geometry ($P_g$, $P_r$, $N_p$) dominates migration angle while DI acts as a contextual modifier through strong interactions (Fig.~\ref{fig:stat_visu}e–f)—we trained supervised regressors to learn the forward map

\begin{equation}
f:\ \{P_g,\,P_r,\,N_p,\,DI\}\ \longrightarrow\ \theta_m
\end{equation}

with the goal of deploying the most accurate surrogate in an inverse design loop. Six models were benchmarked on a held-out test set: XGBoost, Support Vector Regression (SVR), Random Forest (RF), Polynomial Ridge (PR), Elastic Net (EN), and Linear Regression (LR). As summarized in Fig.~\ref{fig:model_visu}a–b and Table~\ref{tab:model_comparison}, XGBoost and SVR attained near-perfect agreement with simulation ground truth ($R^2=0.9999$), with XGBoost exhibiting the lowest test error (MSE $=2\times10^{-4}$). RF performed slightly lower ($R^2=0.9973$), while linear baselines (EN, LR) underfit the nonlinear response ($R^2<0.90$). These results, consistent with the nonlinearity and interaction structure revealed by ANOVA, indicate that tree-based ensembles capture the governing relationships most faithfully.

Figure ~\ref{fig:model_visu}(a–b) Comparative performance across models: panel (a) reports error metrics (e.g., MSE, RMSE, MAE) with variability bars from repeated train/test splits; panel (b) shows $R^2$ distributions. Together, they establish the superiority and stability of XGBoost (and, secondarily, SVR) for this task. 
\textbf{(c)} Parity (predicted vs.\ observed) for XGBoost with the ideal $y{=}x$ line and tolerance bands. Points cluster tightly within $\pm0.1^{\circ}$, demonstrating sub-degree fidelity needed for separation design. 
\textbf{(d)} Residuals versus observation index. The residual cloud is zero-mean with no discernible trend, indicating homoscedastic errors and absence of temporal/ordering artifacts—properties desirable for reliable inversion.
\textbf{(e–f)} Two-dimensional interaction maps (e.g., partial dependence/ALE surfaces) emphasizing \emph{periodicity} couplings: (e) $(N_p,P_g)$ and (f) $(N_p,P_r)$ at representative settings of the remaining variables. The surfaces reveal steep gradients along $N_p$ with curvature induced by $P_g$/$P_r$, mirroring the statistically significant DI–geometry interactions in Fig.~\ref{fig:stat_visu}f. Practically, these maps expose design “levers” and trade-offs: modest adjustments in $P_g$ or $P_r$ can amplify or suppress the $N_p$ effect, enabling targeted tuning of $\theta_m$.

While SVR matches XGBoost in $R^2$, XGBoost attains strictly lower test error and offers fast batched inference with native access to interaction visualizations (Fig.~\ref{fig:model_visu}e–f). We therefore adopt XGBoost as the forward surrogate for inverse design: given target cellular traits (radius and DI), we sweep or optimize over $(P_g, P_r, N_p)$ to maximize angle separation between phenotypes. This choice aligns with the mechanistic picture developed above—geometry sets the baseline, DI modulates through interactions—and provides a high-fidelity, interpretable engine for the optimization results presented next.

\begin{figure}[htbp]
  \centering
  \includegraphics[width=\textwidth]{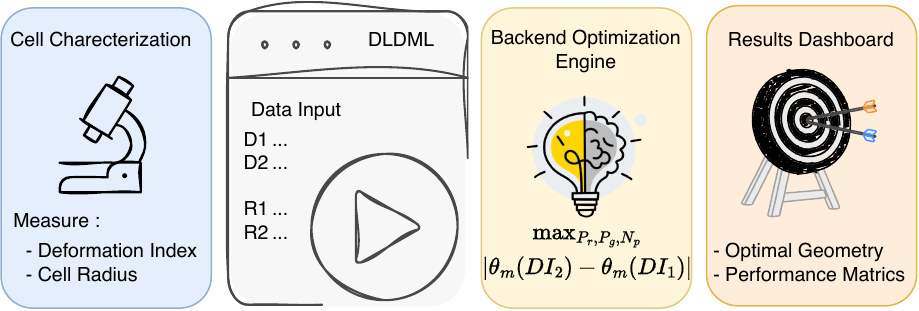}
  \caption{ML-driven DLD design workflow. Experimentally measured DI and cell radii are input to the DLDML web platform, which employs TPE-guided Bayesian optimization with an XGBoost surrogate to maximize inter-population migration angle separation $|\theta_m(\mathrm{DI}_2) - \theta_m(\mathrm{DI}_1)|$ over geometric parameters ($P_r$, $P_g$, $N_p$). The results dashboard provides optimal geometry specifications, performance metrics, and fabrication-ready outputs.}
  \label{fig:workflow}
\end{figure}

\subsection{Design Space Exploration Using Trained ML Surrogates}

Having established XGBoost as the optimal forward predictor mapping $\{\overline{\mathrm{DI}}, P_r, P_g, N_p\} \rightarrow \theta_m$ with sub-degree precision ($R^2 = 0.9999$), we now deploy this surrogate as the computational engine for inverse geometry design. The central objective addressed in this subsection is the converse problem: given a target cellular phenotype characterized by its deformation index, identify the DLD geometry configuration that maximizes inter-population separation performance. This inverse capability transforms the trained surrogate from a passive prediction tool into an active design platform, enabling rapid device optimization without iterative experimental fabrication or exhaustive high-fidelity simulation.

The inverse design task is formulated as a constrained optimization problem where the goal is to maximize the angular separation between two cell populations with distinct mechanical phenotypes. Let $\overline{\mathrm{DI}}_1$ and $\overline{\mathrm{DI}}_2$ denote the measured DI of the two target cell types. The optimal geometry configuration $(P_r^*, P_g^*, N_p^*)$ must satisfy:

\begin{equation}
(P_r^*, P_g^*, N_p^*) = \argmax_{P_r, P_g, N_p} \; \left| f_{\text{XGB}}(P_r, P_g, N_p, \overline{\mathrm{DI}}_1) - f_{\text{XGB}}(P_r, P_g, N_p, \overline{\mathrm{DI}}_2) \right|
\label{eq:inverse_objective}
\end{equation}

subject to manufacturing and operational constraints:
\hspace{.5cm}
\begin{align}
P_r &\in [P_{r,\min}, P_{r,\max}], \label{eq:constraint_radius} \\
P_g &\in [P_{g,\min}, P_{g,\max}], \label{eq:constraint_gap} \\
N_p &\in \mathbb{Z}^+ \cap [N_{p,\min}, N_{p,\max}], \label{eq:constraint_period} \\
P_g &> P_r + d_{\text{safety}}, \label{eq:clogging_constraint}
\end{align}

where $f_{\text{XGB}}$ denotes the trained XGBoost forward model, and the constraints encode feasible fabrication limits for soft lithography or stereolithography (Eqs.~\ref{eq:constraint_radius}--\ref{eq:constraint_period}), as well as minimum gap requirements to prevent particle clogging and maintain hydraulic throughput (Eq.~\ref{eq:clogging_constraint}). To efficiently navigate this constrained, nonlinear design space, we employ Bayesian optimization via Tree-structured Parzen Estimator (TPE)\cite{bergstra2011algorithms}. Unlike gradient-based methods that require differentiability or grid search approaches that scale exponentially with dimensionality, TPE constructs separate probabilistic models for promising and unpromising regions of the parameter space, iteratively refining the search toward high-performing configurations. The TPE algorithm models the conditional probability $p(P_r, P_g, N_p | \Delta\theta)$ using kernel density estimation, splitting observed trials into two groups based on a quantile threshold $\gamma$ and computing an acquisition function that balances exploration of uncertain regions with exploitation of known high-performance zones. Specifically, TPE defines two densities: $\ell(x)$ representing the distribution of parameters $x$ that yielded objective values below a threshold $y^*$ (i.e., good configurations), and $g(x)$ representing parameters above this threshold (i.e., poor configurations). The algorithm then proposes new parameter configurations by maximizing the expected improvement, which is proportional to $\ell(x)/g(x)$, thereby favoring regions where good observations are dense relative to poor ones.

The optimization objective quantifies separation performance through the migration angle difference between the two cell populations. For a given geometry configuration $(P_r, P_g, N_p)$, the XGBoost surrogate predicts migration angles $\theta_1 = f_{\text{XGB}}(P_r, P_g, N_p, \overline{\mathrm{DI}}_1)$ and $\theta_2 = f_{\text{XGB}}(P_r, P_g, N_p, \overline{\mathrm{DI}}_2)$ corresponding to cells with deformation indices $\overline{\mathrm{DI}}_1$ and $\overline{\mathrm{DI}}_2$, respectively. The separation metric is defined as:

\begin{equation}
\Delta\theta(P_r, P_g, N_p) = |\theta_2 - \theta_1|
\label{eq:separation_metric}
\end{equation}

Physical and manufacturing constraints are enforced through penalty functions: geometries violating the gap constraint $P_g \leq P_r$ return a large penalty value ($10^6$) that effectively excludes them from consideration, while all other constraints are satisfied through appropriate parameter bounds defined at the sampler level. The multivariate TPE sampler further enhances optimization efficiency by modeling parameter dependencies. As demonstrated in the preceding ANOVA analysis, DLD performance exhibits strong interaction effects between geometry and deformation index, with statistically significant coupling between $\overline{\mathrm{DI}}$ and geometric parameters ($p < 10^{-30}$ for gap and radius interactions). The multivariate TPE formulation explicitly accounts for these correlations, jointly sampling parameter combinations rather than treating each dimension independently, thereby accelerating convergence to globally optimal solutions.

\subsubsection{Optimization Protocol and Convergence Characteristics}

The optimization protocol proceeds in two phases. An initial exploration phase samples $n_{\text{startup}}$ parameter combinations uniformly at random across the constrained search space, establishing a baseline distribution of separation performance. This startup phase, typically comprising 10--20 trials, ensures adequate coverage of the design manifold before transitioning to model-guided sampling. In the subsequent exploitation phase, the TPE sampler constructs conditional density models from the accumulated trial history and proposes new geometries by maximizing the expected improvement in separation angle. This adaptive strategy balances global exploration of parameter space with local refinement around high-performing regions, converging to near-optimal solutions within $n_{\text{total}} = 50$--200 trials depending on the complexity of the separation landscape.

Convergence is monitored through the best-observed separation angle as a function of trial number. Typically, the optimization exhibits rapid initial improvement during the startup phase as the sampler discovers favorable regions, followed by diminishing returns as subsequent trials refine the geometry through increasingly marginal adjustments. Empirically, we observe that 80--90\% of the achievable separation improvement is realized within the first 50 trials, with later trials contributing fine-tuning adjustments on the order of $0.1^{\circ}$--$0.5^{\circ}$. The computational efficiency of the XGBoost surrogate enables sub-second evaluation of each trial, yielding total optimization times of $10$--$60$ seconds for typical problem instances—a dramatic acceleration compared to the hours or days required for iterative experimental prototyping or direct high-fidelity simulation.

\subsection{Deployment as a Predictive Web Application}

To democratize access to the inverse design framework and eliminate barriers associated with specialized computational infrastructure or machine learning expertise, we developed a full-stack web application comprising a RESTful backend API and an interactive frontend interface\cite{DLDML_github}. The backend, implemented using the FastAPI framework, encapsulates the trained XGBoost surrogate and the TPE optimization engine as microservices, exposing parameterized geometry optimization as a stateless HTTP endpoint. The API accepts JSON-formatted requests specifying target deformation indices ($\overline{\mathrm{DI}}_1$, $\overline{\mathrm{DI}}_2$), cell radii ($R_1$, $R_2$), and constraint bounds on geometric parameters, validates input parameters against physical feasibility criteria, executes the TPE-guided optimization, and returns a comprehensive result payload including optimal geometry, separation angle, convergence history, and parameter importance rankings.

The frontend, provides a zero-code interactive interface for specifying optimization parameters through intuitive sliders and numerical inputs. Upon initiating optimization, the frontend asynchronously polls the backend API and displays real-time progress indicators, enabling users to monitor convergence as the TPE sampler explores the design space. Post-optimization, the interface presents results through a multi-panel dashboard featuring: (i) optimal geometry specifications with corresponding separation angle metrics, (ii) parameter importance bar charts elucidating which design variables most strongly influence performance, (iii) parallel coordinate plots revealing parameter interactions and multi-dimensional relationships, and (iv) interactive contour surfaces allowing users to visualize how separation angle varies across two-dimensional slices of the geometry space. Results are exportable in both JSON and CSV formats, facilitating integration with downstream CAD workflows for mask layout generation or direct import into microfabrication control software.

The application is containerized using Docker to ensure reproducible deployment across heterogeneous computing environments. The multi-container architecture separates the FastAPI backend and Streamlit frontend into independent services orchestrated via Docker Compose, enabling horizontal scaling of the API layer to accommodate concurrent optimization requests without frontend coupling. This modular design further permits cloud deployment on platforms supporting containerized workloads, with the current implementation validated on AWS EC2 instances. The lightweight computational footprint—attributable to the XGBoost surrogate's sub-millisecond inference time—allows execution on minimal cloud infrastructure (single-core, 1~GB RAM instances), substantially lowering the barrier to adoption for resource-constrained laboratories.

\subsubsection{Workflow Integration and Practical Deployment}

The end-to-end workflow from cellular phenotype characterization to fabrication-ready DLD geometry proceeds as follows. Experimentalists first measure the deformation index of target cell populations using standard microfluidic constriction assays, optical stretching, or atomic force microscopy, yielding quantitative DI values with typical measurement uncertainties of $\pm 3$--$5\%$. These measured DI values, along with cell radii and desired operational constraints (e.g., maximum pillar gap to prevent off-target particle escape, minimum gap to avoid clogging), are entered into the web interface. The optimization executes autonomously, typically completing within one minute, and returns a ranked list of candidate geometries. Users select the preferred design based on the balance between predicted separation angle, parameter sensitivity, and manufacturing feasibility for their available fabrication method (soft lithography, 3D printing, or photolithography).

The selected geometry is exported as a structured data file containing pillar radius, gap dimensions, and periodicity specifications. These parameters serve as direct inputs to mask design software for photolithography-based fabrication or are translated into G-code for stereolithography 3D printing. For soft lithography workflows, the geometry specifications define the master mold dimensions, which are subsequently replicated in polydimethylsiloxane (PDMS) to yield the functional DLD device. Post-fabrication validation involves optical microscopy to verify dimensional accuracy and test separations with calibration particles or cell lines of known deformability to confirm predicted performance. Discrepancies between predicted and observed separation angles, typically arising from fabrication tolerances or flow regime deviations, can be addressed through iterative refinement: measured performance data are fed back into the optimization framework with adjusted constraints, converging to experimentally validated designs within two to three fabrication-test cycles.

The integration of high-fidelity simulation-trained surrogates with Bayesian optimization and accessible web deployment addresses the historical bottleneck in DLD device development—the disconnect between computational prediction and experimental realization. By reducing the design iteration cycle from weeks of manual experimentation to hours of automated optimization followed by single-pass fabrication, this framework accelerates the translation of DLD technology from research-grade prototypes to application-specific, clinically relevant devices. Moreover, the open architecture of the web application permits community contributions: researchers can retrain surrogates on expanded simulation datasets encompassing additional geometric degrees of freedom (e.g., pillar shape, surface chemistry) or alternative cell types, uploading updated models to extend the platform's predictive scope without modifying the underlying optimization infrastructure.

\section*{Conclusion}

This study establishes a physics-guided machine learning framework that transforms DLD device design from an empirical, trial-and-error process into a data-driven, inverse optimization workflow. By integrating high-fidelity LB and IB simulations with gradient-boosted surrogate models, we have systematically resolved the coupling between cellular mechanical phenotype and microfluidic array geometry. Our results demonstrate that while membrane stiffness properties (stretching modulus $K_s$ and area-dilation modulus $K_{\alpha}$) govern the DI with correlation coefficients exceeding 0.7, geometric parameters (pillar gap $P_g$, pillar radius $P_r$, and periodicity $N_p$) dominate migration angle variance through statistically significant interactions (Type II ANOVA, $p < 10^{-30}$). The XGBoost surrogate model achieves near-perfect predictive accuracy ($R^2 = 0.9999$, MSE $= 2 \times 10^{-4}$), enabling Bayesian optimization via TPE to identify fabrication-ready geometries in under one minute. The web-deployed framework reduces the design-to-fabrication cycle from weeks of iterative experimentation to hours of automated optimization followed by single-pass prototyping, thereby democratizing access to optimized DLD systems for resource-constrained laboratories. The framework presented here addresses a critical gap in the DLD literature by delivering not merely forward predictions but actionable inverse designs with explicit robustness quantification. By converting measured DI into turnkey device specifications, this pipeline bridges cellular biomechanics and microfabrication engineering in a manner that is both computationally rigorous and experimentally accessible. The computational validation through cross-validated surrogate models ($R^2 > 0.99$) and convergent Bayesian optimization establishes the internal consistency of the framework, while the identification of robust geometric regions provides design guidelines that inherently account for manufacturing variability. Future extensions will broaden the applicability of this methodology in three directions. First, the simulation framework can be extended to incorporate non-Newtonian fluid rheology and viscoelastic effects to model complex biological fluids such as blood and synovial fluid. Second, three-dimensional pillar geometries (e.g., cylindrical, triangular, or I-shaped cross-sections) and multi-stage cascaded arrays can be integrated to address high-throughput clinical applications requiring processing rates exceeding $10^6$ cells per hour. Third, coupling with real-time imaging and machine vision systems would enable adaptive, closed-loop geometry tuning in response to heterogeneous or evolving sample populations. These extensions, combined with the modular architecture of the web platform, position machine learning-augmented DLD design as a scalable and adaptable methodology for label-free cell separation across oncology, prenatal diagnostics, and regenerative medicine.

\bibliography{main}

\end{document}